\documentclass[aps,
twocolumn,
showpacs,
superscriptaddress,groupedaddress]{revtex4}

\usepackage{amssymb}
\usepackage{amsmath}
\usepackage{float}
\usepackage{graphicx}
\usepackage[caption=false]{subfig}
\usepackage{subfig}
\usepackage{mathrsfs}

\begin{document}

\title{Theory of the crossover from lasing to steady state superradiance}

\author{D. A. Tieri} 

\affiliation{JILA and Department of Physics, University of Colorado,
  Boulder, Colorado 80309-0440, USA.}

\author{Minghui Xu} 

\affiliation{Department of Physics and Astronomy, Shanghai Jiao Tong
  University, Shanghai 200240, China.}

\author{D. Meiser} 

\altaffiliation{present address: Trimble Boulder, 4730 Walnut Street,
  Suite 201, Boulder, CO 80301, USA.}

\affiliation{Tech-X Corporation, 5621 Arapahoe Avenue, Boulder,
  Colorado 80303, USA.}

\author{J. Cooper}

\affiliation{JILA and Department of Physics, University of Colorado,
  Boulder, Colorado 80309-0440, USA.}

\author{M. J. Holland} 

\affiliation{JILA and Department of Physics, University of Colorado,
  Boulder, Colorado 80309-0440, USA.}

\date{\today}

\begin{abstract}
  Lasing and steady state superradiance are two phenomena that may
  appear at first glance to be distinct.  In a laser, phase
  information is maintained by a macroscopic intracavity light field,
  and the robustness of this phase is what leads to the coherence of
  the output light.  In contrast, the coherence of steady-state
  superradiant systems derives from the macroscopic collective dipole
  of a many-atom ensemble.  In this paper, we develop a quantum theory
  that connects smoothly between these two extreme limits.  We show
  that lasing and steady-state superradiance should be thought of as
  the two extreme limits of a continuous crossover. The properties of
  systems that lie in the superradiance, lasing, and crossover
  parameter regions are compared.  We find that for a given output
  intensity a narrower linewidth can be obtained by operating closer
  to the superradiance side of the crossover.  We also find that the
  collective phase is robust against cavity frequency fluctuations in
  the superradiant regime and against atomic level fluctuations in the
  lasing regime.
\end{abstract}


\maketitle

\section{Introduction}

Since its first demonstration in 1960~\cite{maiman1960stimulated}, the
laser has had a profound impact on fundamental science research and has
found widespread applications in society in general.  Although many
different types of lasers exist, with their characteristic parameters
(such as power, linewidth, pulse duration, and physical size) spanning
many orders of magnitude, all lasers share a common conceptual
foundation.  A laser is a cavity quantum electrodynamics (QED) system
consisting of a gain medium inside an optical
cavity~\cite{meystre2007elements}.  We will oftentimes refer to the gain
medium as ``atoms'' for brevity.  Lasers typically operate in the good
cavity regime of cavity QED where the linewidth of the cavity is much
narrower than the bandwidth of the gain medium.  The atoms generate a
coherent electromagnetic field in the cavity by means of stimulated
emission~\cite{PhysRev.112.1940}.  Stimulated emission is a quantum
mechanical interference effect in which the presence of a large number
of photons in a particular mode of a light field increases the
probability that an atom will emit into that mode. In a laser, the
macroscopic phase information that is associated with the coherence of
the generated radiation is encoded in the light field.

Around the same time as the laser was first demonstrated, the effect
of superradiance was predicted~\cite{PhysRev.93.99}, and soon
thereafter experimentally demonstrated~\cite{harocheSuperradiance}.
Superradiance is a quantum mechanical interference effect in which
correlations between atoms lead to collective emission.  Superradiance
has most commonly been considered as a transient phenomenon.  Atoms in
an ensemble are prepared in the excited state.  Spontaneous emission
is then enhanced via the growth of atom-atom correlations.  However,
it has been known for some time that superradiance can also occur in
steady state~\cite{PhysRevLett.102.163601, PhysRevA.81.033847,
  PhysRevA.81.063827,PhysRevLett.89.253003} by placing the atomic
ensemble inside a cavity.  In contrast to lasers, superradiance in
steady-state occurs in a cavity with a much broader linewidth than the
atomic linewidth.  This regime is referred to as the bad-cavity limit
of cavity QED~\cite{PhysRevA.51.809, PhysRevLett.72.3815,
  ChenDeliciousLaser, HakenLaser, HakenLaserBook}.  The radiation
produced in steady state superradiance is also coherent.  However, in
contrast to a laser, the coherence is encoded in the atomic medium.
Progress has recently been made towards the experimental realization
of steady-state superradiant
systems~\cite{ThompsonPaper,bohnet2012relaxation,Norcia:Crossover}.

An important application of lasers is as a stable local oscillator for
optical atomic clocks and precision
spectroscopy~\cite{RevModPhys.87.637}.  These lasers rely on
stabilization against reference cavities.  The most advanced such
lasers reach linewidths below $0.1 {\rm Hz}$ corresponding to quality
factors of $Q>10^{15}$~\cite{Cole:TenfoldReductionBrownianNoise}. The
principal limiting factor in the way of further improvement of these
local oscillators is thermal vibrations of the dielectric coatings on
the cavity mirrors~\cite{PhysRevLett.101.260602}.  To overcome this
technical challenge, researchers have proposed an alternate approach
using an active system based on steady state superradiance on a clock
transition to create an even more stable light
source~\cite{PhysRevLett.102.163601, ChenDeliciousLaser}.  However,
this proposal has challenges of its own.  First of all, in spite of
the enhancement that occurs due to superradiance, the produced
intensity is orders of magnitude lower than for a conventional laser.
Second of all, perturbations of atomic transition frequencies can
potentially lead to phase and frequency perturbations in the generated
field.

In this paper we develop a unified theory of lasers and steady state
superradiance.  We show that lasers and steady state superradiance are
the extreme limits of a continuous crossover.  The theory allows us to
directly compare and contrast lasers, steady state superradiant
systems, and systems in the crossover region using common language.
Our analysis further clarifies the qualitative and quantitative
differences between lasing and steady state superradiance.  From the
perspective of applications, the unified theory enables us to
determine the optimal system for ultra-stable local oscillators and
precision measurement applications.

We analyze the model using different levels of approximation: an exact
method using Monte-Carlo trajectories and $\mathrm{SU}(4)$ operators, a
semi-classical method based on \textit{c}-number Langevin equations, a
quantum phase diffusion model, and a mean-field model.  The different
approaches provide insight into different aspects of the problem.
Highly simplified models like the mean field equations and phase
diffusion yield a qualitative understanding of the general
characteristics of systems throughout the crossover.  By comparison
between the approximations we can differentiate between truly critical
physical effects and less important details.  We find that fluctuations
and correlations play an important role in the the noise properties of
the system (e.g.\ the linewidth of the generated light), but can be
modelled semi-classically.  Comparison with the exact $\mathrm{SU}(4)$
method for small numbers of atoms shows that \textit{c}-number Langevin
equations provide an accurate description of the system.  Due to their
much smaller computational complexity, we are then able to use the
\textit{c}-number Langevin equations to quantitatively study much larger
systems relevant for experiment.

The rest of this paper is organized as follows.  In
Section~\ref{sec:Model} we summarize the physical model upon which our
analysis is based. In Section~\ref{sec:Methods} we discuss several
approximation methods.  We compare the approximations with one another
to determine their accuracy and to evaluate their ability to capture the
various physical signatures. In
Section~\ref{sec:CrossoverCharacterization} we define a crossover
parameter which characterizes the relative importance of stimulated
emission to collective atomic effects in a cavity QED system.  In
Section~\ref{sec:Results} we discuss our results on the crossover.

\section{Model}
\label{sec:Model}

As noted in the introduction, the fundamental ingredients of lasers
and superradiance systems are an electromagnetic field and atoms
serving as a gain medium.  A minimal model consists of a single mode
cavity field and an ensemble of $N$-two level atoms.  The atoms couple
to the cavity field via the dipole interaction.  Energy is supplied to
the system by means of incoherent repumping mechanisms.  In practice
this necessitates auxiliary atomic levels that rapidly decay and can
be adiabtically eliminated. The resulting effect is incoherent
transfer of population from the ground to excited state. The
incoherent repumping, along with the atomic spontaneous emission,
cavity decay, and other relaxation processes, make the system
fundamentally an open quantum system that requires a quantum master
equation treatment.

Mathematically, our model is described by the quantum master equation
derived in the Born and Markov approximations for the reduced density
matrix of the system $\hat{\rho}$,
\begin{equation}
  \frac{d}{dt} \hat{\rho} =
  \frac{1}{i \hbar} \left[ \hat{H}, \hat{\rho} \right] +
  \hat{\mathcal{L}}\left[ \hat{\rho} \right],
\label{ME1Crossover}
\end{equation}
where,
\begin{equation}
\hat{H} = \frac{\hbar \omega_a}{2} \sum_{j=1}^{N} \hat{\sigma}^{z}_{j}
+ \hbar \omega_c \hat{a}^{\dagger}\hat{a}
+ \frac{\hbar \Omega}{2}  \sum_{j=1}^{N} \left(
    \hat{a}^{\dagger} \hat{\sigma}^{-}_{j} +
    \hat{\sigma}^{+}_{j} \hat{a} \right)\;.
\end{equation}
The Hamiltonian $\hat{H}$ describes the coherent evolution of the
coupled atom cavity system, where $\omega_{a}$ is the atomic
transition frequency and $\omega_c$ is the frequency of the cavity
mode. The Pauli spin matrices for the $j$-th atom are
$\hat{\sigma}_j^{+}$, $\hat{\sigma}_j^{-}$ and $\hat{\sigma}_j^{z}$,
and $\hat{a}$ is the annihilation operator of the cavity mode. The
atom-cavity coupling strength is~$\Omega$.  In general, the
atom-cavity coupling depends on the location of the atom in the cavity
field.  To simplify the discussion, we ignore the spatial dependence
since it results in quantitative changes but does not alter the basic
physical properties.  In principle, a constant $\Omega$ could be
realized experimentally by confining the atoms to locations of equal
amplitude of the cavity mode by means of a superimposed optical
lattice.

The Liouvillian superoperator
$\hat{\mathcal{L}}\left[ \hat{\rho} \right]$ describes the various
non-Hermitian processes,
\begin{eqnarray}
\hat{\mathcal{L}}\left[ \hat{\rho} \right] &=&
  -\frac{\kappa}{2}
  \left(
    \hat{a}^{\dagger} \hat{a} \hat{\rho}
    + \hat{\rho}  \hat{a}^{\dagger} \hat{a}
    - 2\hat{a} \hat{\rho} \hat{a}^{\dagger}
  \right)
\nonumber
\\
 &&-\frac{\gamma}{2} \sum_{j=1}^N
  \left(
   \hat{\sigma}_{j}^{+} \hat{\sigma}_{j}^{-} \hat{\rho}
   + \hat{\rho} \hat{\sigma}_{j}^{+} \hat{\sigma}_{j}^{-}
   - 2\hat{\sigma}_{j}^{-} \hat{\rho} \hat{\sigma}_{j}^{+}
  \right)
\nonumber
\\
 &&-\frac{w}{2} \sum_{j=1}^N
  \left(
   \hat{\sigma}_{j}^{-} \hat{\sigma}_{j}^{+} \hat{\rho}
   + \hat{\rho} \hat{\sigma}_{j}^{-} \hat{\sigma}_{j}^{+}
   - 2\hat{\sigma}_{j}^{+} \hat{\rho}  \hat{\sigma}_{j}^{-}
  \right),
\nonumber
\\
 &&+\frac{1}{2T_2} \sum_{j=1}^N
  \left(
   \hat{\sigma}_{j}^{z} \hat{\rho}  \hat{\sigma}_{j}^{z} - \hat{\rho}
  \right),
\end{eqnarray}
where $\kappa$ is the decay rate of the cavity, $\gamma$ is the
free-space spontaneous emission rate of the atoms, $w$ is the
repumping rate, and $1/{T_2}$ is the rate of inhomogeneous dephasing.

\section{Solution Methods}
\label{sec:Methods}

Obtaining a direct numerical solution to Eq.~(\ref{ME1Crossover}) is
impossible for experimentally relevant numbers of particles because the
dimension of the Hilbert space of the system scales as $2^N$.  In this
Section we introduce several solution methods to overcome the
exponential scaling of the size of the Hilbert space.  They can be
grouped into three categories: Exact methods ($\mathrm{SU}(4)$ method
with Monte-Carlo simulation), semi-classical methods (\textit{c}-number
Langevin equations, phase diffusion), and mean-field treatments.

The exact solution methods solve the quantum mechanical problem directly
without further approximations, but are limited in applicability to
small numbers of atoms. The $\mathrm{SU}(4)$ method provides an exact
numerical solution of Eq.~(\ref{ME1Crossover}) by exploiting an
underlying permutation symmetry to drastically reduce the Hilbert space
dimension~\cite{Hartmann:arXiv1201.1732, PhysRevA.87.062101}. Details of
the method have been described previously in~\cite{PhysRevA.87.062101}.
Here we extend the approach to solve the quantum master equation in the
$\mathrm{SU}(4)$ representation using the quantum jump
method~\cite{Dalibard92,Dum92,Knight98}.  We give details of the
$\mathrm{SU}(4)$ quantum jump method in Appendix~\ref{Su4Appendix}.

Semi-classical methods aim to capture the physics of the system
correctly for large atom number.  They include a classical
representation of fluctuations and correlations.  Comparison with
direct solution methods for small atom number allows us to verify the
validity and accuracy of the semi-classical approaches.

The mean-field methods neglect fluctuations to arrive at equations for
averaged quantities. These equations are sufficiently simple that it is
straightforward to obtain closed form solutions that provide valuable
qualitative insights into the system behavior.

\subsection{Quantum Langevin Equations}

For the derivation of the semi-classical equations corresponding to
Eq.~(\ref{ME1Crossover}) it is convenient to work in the Heisenberg
picture.  The resulting equations are the quantum Langevin equations
\begin{eqnarray}
\frac{d}{dt} \hat{a}&=& -\frac{1}{2} (\kappa +2i\omega_c) \hat{a}
-\frac{i N \Omega}{2} \hat{S}^{-}
+\hat{F}^{a},
\label{La}\\
\frac{d}{dt} \hat{S}^{-} &=&
-\frac{1}{2} \left(\Gamma +2 i \omega_a \right)  \hat{S}^{-}
+\frac{i \Omega}{2} \hat{a} \hat{S}^{z}
+\hat{F}^{-},
\label{Lsm}\\
\frac{d}{dt} \hat{S}^{z} &=&
-(w+\gamma)\left( \hat{S}^{z} - d_0\right)
+i\Omega \left( \hat{a}^{\dagger}\hat{S}^{-} -
\hat{a}\hat{S}^{+} \right)
+\hat{F}^{z},\nonumber\\
\label{Lsz}
\end{eqnarray}
where $\delta=\omega_{a}-\omega_{c}$ is the atom-cavity detuning,
$\Gamma \equiv w+\gamma+2/T_2$ is the generalized single-atom
decoherence, and $d_0 = (w-\gamma)/(w+\gamma)$ characterizes the
atomic inversion that would be obtained for a single-atom in the
absence of the cavity. We have defined the collective operators,
\begin{eqnarray}
\hat{S}^{\pm}&=&\frac{1}{N}\sum_{k=1}^N \hat{\sigma}_k^{\pm},
\\
\hat{S}^{z}&=&\frac{1}{N}\sum_{k=1}^N \hat{\sigma}_k^{z}\,.
\end{eqnarray}
The quantum noise operators $\hat F^\mu$ have zero mean and
second-order correlations given by
\begin{equation}
\left< \hat{F}^{\mu}(t) \hat{F}^{\nu}(t^{\prime})\right> =
2 D^{\mu \nu} \delta(t-t^{\prime})\;.
\end{equation}
The diffusion matrix elements $D^{\mu \nu} $ are obtained using the
Einstein relations \cite{meystre2007elements},
\begin{eqnarray}
2D^{a a^{\dagger}}&=& \kappa \\
2D^{+-}&=& \frac{1}{N}
\left(
  w + \frac{1}{T_2} \left(1 + \left< \hat{S}^{z} \right> \right)
\right) \\
2D^{-+}&=& \frac{1}{N}
\left(
  \gamma + \frac{1}{T_2} \left(1- \left< \hat{S}^{z} \right> \right)
\right) \\
2D^{+z}&=& -\frac{2w}{N} \left< \hat{S}^{+} \right>
\\
2D^{z+}&=& \frac{2\gamma}{N} \left< \hat{S}^{+} \right>
\\
2D^{-z}&=& \frac{2\gamma}{N} \left< \hat{S}^{-} \right>
\\
2D^{z-}&=& -\frac{2w}{N} \left< \hat{S}^{-} \right>
\\
2D^{zz}&=& \frac{2\gamma}{N}
\left(1+ \left< \hat{S}^{z} \right> \right) +
\frac{2w}{N}\left(1- \left< \hat{S}^{z} \right> \right).
\label{OpNoise1}
\end{eqnarray}

\subsection{\textit{C}-number Langevin Equations}

Quantum Langevin equations are operator valued stochastic differential
equations.  As such they are difficult to numerically simulate. To
obtain numerically tractable equations we construct a semi-classical
theory by replacing the operators in the quantum Langevin equations by
complex numbers,
\begin{eqnarray}
\frac{d}{dt} a&=& -\frac{1}{2}  (\kappa +2i\omega_c) a
-\frac{i N \Omega}{2} S^{-}
+F^{a},
\label{Lac}\\
\frac{d}{dt} S^{-} &=& -\frac{1}{2}  \left(\Gamma +2 i \omega_a \right)  S^{-}
+\frac{i \Omega}{2} a S^{z}
+F^{-},\\
\frac{d}{dt} S^{z} &=& -(w+\gamma)\left( S^{z} - d_0\right)
+i\Omega \left( a^{*}S^{-} - a S^{+} \right)
+F^{z},\nonumber\\
\label{Lszc}
\end{eqnarray}
where there are no hats over the variables to signify that they are
\textit{c}-numbers and not operators.  The noise terms $F^a$, $F^-$, and
$F^z$ should be interpreted according to the rules of Ito calculus. It
is easier to construct the semi-classical equations by introducing real
variables according to
\begin{equation}
\begin{array}[b]{rclrcl}
q &=&
\frac{1}{2} \left( a^{*} + a \right),&
p &=&
\frac{1}{2i} \left( a^{*} - a \right),
\\[1pc]
S^x &=&
\frac{1}{2} \left( S^{+} + S^{-} \right),&
S^y &=&
\frac{1}{2i} \left( S^{+} - S^{-} \right)\,.
\end{array}
\end{equation}
The equations of motion in terms of these variables are
\begin{eqnarray}
\frac{d}{dt} q &=& -\kappa q - 2 \omega_c p - N \Omega S^{y} + F^{q},
\label{cq1}
\\
\frac{d}{dt} p&=& -\kappa p + 2 \omega_c q + N \Omega S^{x} + F^{p},
\\
\frac{d}{dt} S^{x} &=&
-\Gamma S^{x}  - 2 \omega_a S^{y} + \Omega p S^{z} + F^{x},
\\
\frac{d}{dt} S^{y} &=&
-\Gamma S^{y}  + 2 \omega_a S^{x} - \Omega q S^{z} + F^{y},
\\
\frac{d}{dt} S^{z} &=& -(w+\gamma)\left( S^{z} - d_0\right)
+2 \Omega \left( q S^{y} - p S^{x} \right)
+F^{z}\;.
\nonumber
\\
\label{eqn:cnumberlangevin}
\end{eqnarray}
The noise terms have zero mean and delta-correlations given by
\begin{equation}
\left< F^{\mu}(t) F^{\nu}(t^{\prime})\right> =
2 \mathscr{D}^{\mu \nu} \delta(t-t^{\prime})\;.
\label{ClassicalDiffusion1}
\end{equation}
The correspondence between the semi-classical and quantum mechanical
Langevin equations is established by requiring that they produce
identical equations for first and second moments of the system
operators.  Comparison of the second moments allows us to find the
classical diffusion matrix elements $\mathscr{D}^{\mu \nu}$.  In order
to make this procedure well defined we have to choose a specific
ordering of the quantum mechanical operators.  We choose to make the
correspondence using symmetric ordering defined by the symmetric
expectation value
\begin{equation}
\left< \hat{A}^{\mu} \hat{A}^{\nu} \right>_s=
\frac{1}{2} \left( \left< \hat{A}^{\mu} \hat{A}^{\nu} \right> + \left<
\hat{A}^{\nu} \hat{A}^{\mu} \right> \right)\;,
\end{equation}
where $\hat{A}^{\mu}$ and $\hat{A}^{\nu}$ are system operators.  We
point out that in this formulation, the classical Langevin equations are
equivalent to a Fokker-Planck equation for the Wigner quasi-probability
distribution. The resulting diffusion matrix elements are
\begin{eqnarray}
  2\mathscr{D}^{q q}&=&
                        2\mathscr{D}^{p p}=
                        \frac{\kappa}{4} \nonumber \\
  2\mathscr{D}^{xx}&=&
                       2\mathscr{D}^{yy}=
                       \frac{\Gamma}{4N} \nonumber \\
  2\mathscr{D}^{xz}&=&
                       2\mathscr{D}^{zx}=
                       \frac{-w+\gamma}{N} \left< S^{x} \right>  
                       \nonumber \\
  2\mathscr{D}^{yz}&=&
                       2\mathscr{D}^{zy}=
                       \frac{-w+\gamma}{N} \left< S^{y} \right>  
                       \nonumber \\
  2\mathscr{D}^{zz}&=&
                       \frac{2}{N}\left((w+\gamma) 
                       + (-w+\gamma)  \left< S^{z} \right> \right)\;.
\label{cNoise1}
\end{eqnarray}

We solve the stochastic differential equations,
Eqs.~(\ref{cq1})--(\ref{eqn:cnumberlangevin}) by means of an explicit
second order weak scheme~\cite{kloeden2011numerical}. We find
empirically that the symmetrically ordered diffusion matrix is positive
definite when the system is above the first threshold (defined in
Sec.~(\ref{MFE})). Below this threshold, the symmetrically ordered
diffusion matrix is not positive definite, and divergent trajectories
can occur. We numerically evolve an ensemble of trajectories
simultaneously and we compute the expectation values appearing in
Eq.~(\ref{cNoise1}) as ensemble averages. This allows us to use the
additive form of the explicit second order weak scheme, which is simpler
to implement than the general form. Typically, an ensemble of 1000
trajectories is sufficient to achieve convergence to within a few
percent.

A specialization of the \textit{c}-number Langevin approach that includes
fluctuations in the phase (but not amplitude) of the photon field has
been presented by Haken~\cite{HakenLaserBook}, and provides a
closed-form solution to obtain the spectral linewidth of the output
field. We will refer to this as the phase diffusion method, and the
details are given in Appendix~\ref{HakenAppendix}.

\subsection{Mean-Field Treatment}
\label{MFE}

The mean field equations capture many of the most important features of
the physical system because the noise terms scale in general as
$\sqrt{N}$ while the expectation values scale as~$N$.  In the limit of
large numbers of atoms the noise terms are therefore typically less
important for certain quantities.

By taking expectation values of the semi-classical
Eqs.~(\ref{cq1}--\ref{eqn:cnumberlangevin}) we obtain mean-field
equations written in the reference frame rotating at frequency
$\omega$
\begin{eqnarray}
\frac{d}{dt} a_0&=& -\frac{1}{2} (\kappa +2i(\omega_c-\omega)) a_0
-\frac{i N \Omega}{2} S_0^{-}\;,
\label{La0}\\
\frac{d}{dt} S_0^{-} &=&
-\frac{1}{2} \left(\Gamma +2 i (\omega_a-\omega) \right)  S_0^{-}
+\frac{i \Omega}{2} a_0 S_0^{z}\;,\\
\frac{d}{dt} S_0^{z} &=& -(w+\gamma)\left( S_0^{z} - d_0\right)
+i\Omega \left( a_0^{*} S_0^{-} - a_0 S_0^{+} \right)\;,\nonumber\\
\label{Lsz0}
\end{eqnarray}
where the $0$ subscript denotes the mean value, {\it e.g.}\ $\langle
\hat{a} \rangle =a_0$.  Noise terms do not appear since they have zero
average.

A closed-form solution of Eqs.~(\ref{La0})--(\ref{Lsz0}) can be
obtained in steady-state by setting the left hand sides to zero.  We
find
\begin{equation}
S_0^{z}=
\frac{(\kappa+2i(\omega_c-\omega))(\Gamma+2i(\omega_a-\omega))}{N\Omega^2}
\label{Sz01}
\end{equation}
for the steady state inversion. The oscillation frequency of the
atom-cavity coupled system $w$ can be determined using the condition
that $S_0^{z}$ must be real, giving
\begin{equation}
\omega = \frac{\kappa \omega_a + \Gamma \omega_c}{\kappa+\Gamma}\;.
\label{atomcavityfrequencycenter1}
\end{equation}
Simple expressions for atomic inversion and intracavity photon number
can be obtained in the limit of
$\delta = \omega_a-\omega_c \ll \Gamma,\kappa$. We find
\begin{eqnarray}
  S_0^{z}&\approx& \frac{1}{\mathcal{C}}\nonumber\\
  |a_0|^2&\approx&\frac{N(w+\gamma)}{2 \kappa}
             \left(d_0 - \frac{1}{\mathcal{C}}\right)\;,
\label{a0sqSS}
\end{eqnarray}
where $\mathcal{C}\equiv \frac{N \Omega^2}{\kappa \Gamma}$ is the
generalized many-atom cooperativity parameter.  We refer to this as a
generalized parameter since the cooperativity is typically defined in
terms of the single-atom linewidth $\gamma$, but here the effective
linewidth $\Gamma$ includes the dephasing $1/T_2$ and incoherent
repumping $w$ as well.

The zeros of the intra cavity photon number Eq.~(\ref{a0sqSS}) determine
where the system reaches threshold.  The first threshold is obtained at
\begin{equation}
w_1 = \gamma\;,
\label{FirstThreshold}
\end{equation}
which corresponds to the condition that energy must be supplied to the
system at a rate sufficient to maintain population inversion of the
atoms.  A coherent macroscopic field in the cavity emerges and is
accompanied by the formulation of a collective atomic dipole.  A second
threshold occurs at a higher rate of pumping,
\begin{equation}
w_2 =  \frac{N \Omega^2}{\kappa}\;,
\end{equation} 
where we have assumed the collective decay rate $\mathcal{C}\Gamma$ is
much larger than the single atom rates $\gamma$ and $1/T_2$.  The second
threshold corresponds to the situation where the repumping is so strong
that $S_0^{z}$ is close to unity, and the noise due to the strong
repumping prevents the formation of both a macroscopic photon field in
the cavity and a macroscopic dipole in the atomic ensemble.

The photon number in the cavity reaches its maximum at an approximate
repumping strength of
\begin{equation}
  w=w_{\mathrm{opt}}= \frac{N \Omega^2}{2\kappa} - \gamma - \frac{1}{T_2}.
\label{wopt}
\end{equation}
Again assuming the collective decay rate $\mathcal{C}\Gamma$ is much
larger than the single atom rates $\gamma$ and $1/T_2$, we find a
simple expression for the maximum photon number,
\begin{equation}
{(|a_0|^2)}_{\mathrm{opt}}= \frac{N^2 \Omega^2}{8\kappa^2}\;.
\label{adaopt}
\end{equation}

\begin{figure*}[t]
\begin{center}
\includegraphics[scale =0.55] {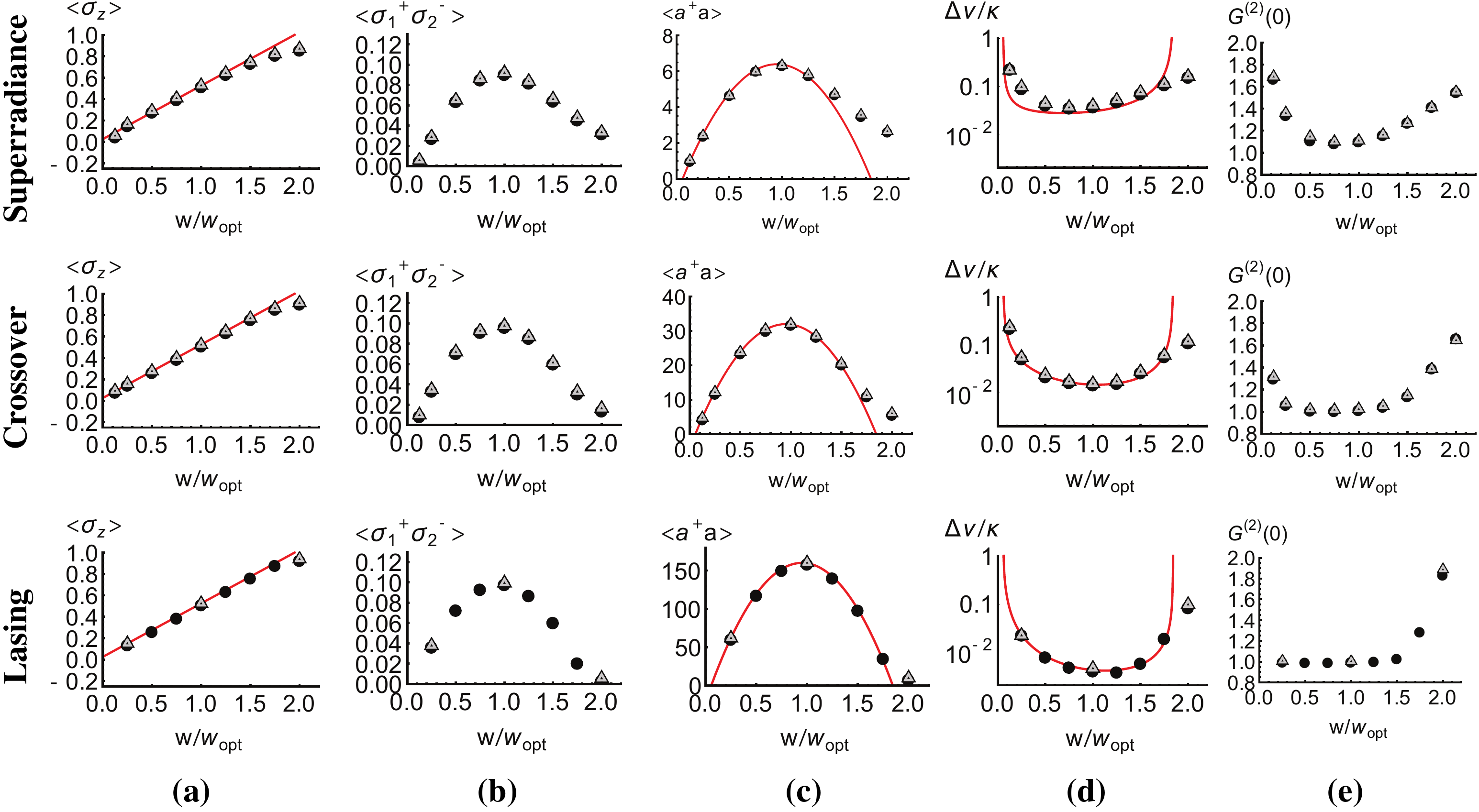}
\end{center}
		\vspace{-5mm}
\caption{(Color online) Comparison of the different solution methods in
the superradiance ($\xi=0.2$), crossover ($\xi=1$), and lasing ($\xi=5$)
regions for $N=40$ and $\frac{\Omega^2}{\kappa \gamma}=1$.  The analytic
Langevin (phase diffusion and mean field) solutions are shown in red, 
the exact $\mathrm{SU}(4)$ solution is shown by grey triangles, and the
\textit{c}-number Langevin simulation results are shown by black
circles. The observables considered are (a) the inversion
$\left<\hat{\sigma}^{z}\right>$, (b) the correlation between the atoms'
dipoles $\left<\hat{\sigma}_{1}^{+} \hat{\sigma}_{2}^{-}\right>$, (c)
the intracavity photon number $\left<\hat{a}^{\dagger}\hat{a}\right>$,
(d) the linewidth $\Delta \nu$, and (e) the intensity correlation
function $G^{(2)}(0)$.}
\label{N40Comparison}
\end{figure*}

\section{Characterization of the Crossover}
\label{sec:CrossoverCharacterization}

The crossover from superradiance to lasing is characterized by a
transition from coherence encoded in an atomic ensemble to coherence
encoded in the light field. The key parameter in identifying the
regime is the ratio of the photon number to atom number. With this
motivation, we introduce a crossover parameter as
\begin{equation}
\xi \equiv \frac{{(|a_0|^2)}_{\rm opt}}{N}\;,
\label{crossoverparam}
\end{equation}
that is, the dimensionless ratio of the maximum intracavity photon
number to the number of atoms. The parameter $\xi$ quantifies the
relative importance of stimulated emission to collective atomic
spontaneous emission. If $\xi\ll1$, the system is in the bad cavity or
superradiant regime. If $\xi\gg1$ the system is in the good cavity or
laser regime. In the crossover or intermediate region, $\xi\sim1$, the
system possesses features of both.

The mean field equations allow us to rewrite this expression in an
alternate way that illuminates the role of the system parameters in
determining the crossover regime. Using Eq.~(\ref{adaopt}), we can
rewrite Eq.~(\ref{crossoverparam}) as
\begin{equation}
  \xi= \frac{N \Omega^2}{8\kappa^2}\;.
\label{CrossoverParameter2}
\end{equation}
The interpretation of this is that the crossover is also characterized
by the ratio of the collective coupling between the many atom ensemble
and the photon mode, $\sqrt{N}\;\Omega$, to the linewidth of the
cavity, $\kappa$.

\section{Results}
\label{sec:Results}

In this section, we present results throughout the crossover from lasing
to superradiance for the field intensity and linewidth, and for the
atomic inversion and correlations. We begin by comparing different
solution methods for small atom numbers.  This comparison shows that the
\textit{c}-number theory gives an accurate description of first and second
moments of the system operators.  With the validity of the semiclassical
method established we then apply it to experimentally relevant systems
with large atom number.

\subsection{Comparison of Different Solution Methods}

In order to determine the validity of the approximate solution methods,
we begin by comparing to the exact $\mathrm{SU}(4)$ Monte-Carlo
simulation for $N=40$ atoms. This is small enough to still be tractable
by exact $\mathrm{SU}(4)$ Monte-Carlo simulations and at the same time
it is large enough to expect the approximate solution methods to be
reasonably accurate.

Fig.~\ref{N40Comparison} shows several observables obtained using the
mean field Langevin method, the phase diffusion method, the
\textit{c}-number Langevin method, and exact $\mathrm{SU}(4)$
Monte-Carlo simulations for three different values of the crossover
parameter: $\xi=0.2$, $\xi=1$, and $\xi=5$. These values of $\xi$ place
the system in the superradiance, crossover, and lasing parameter
regions, respectively.

Figs.~\ref{N40Comparison} (a) and (c) show that the mean field equations
are accurate near the peak of the intracavity photon number, $w=w_{\rm
opt}$, but they are less accurate outside that region.
Fig.~\ref{N40Comparison} (d) shows that the phase diffusion model for
the linewidth also agrees with the exact solution in the region around
$w=w_{\rm opt}$, but disagrees outside that region, where the phase
diffusion approximation breaks down.  Although they do not
quantitatively agree with the exact $\mathrm{SU}(4)$ method, the
analytic solutions obtained by the mean field and phase diffusion models
capture the correct qualitative behavior of the system.

Fig.~\ref{N40Comparison} shows excellent agreement between
\textit{c}-number Langevin and the exact $\mathrm{SU}(4)$ theory in all
parameter regions for all of the considered observables.  Therefore, the
\textit{c}-number Langevin theory can be relied upon for larger atom
numbers inaccessible to the exact numerical solution.

\begin{figure*}
\begin{center}
\includegraphics[scale =0.55] {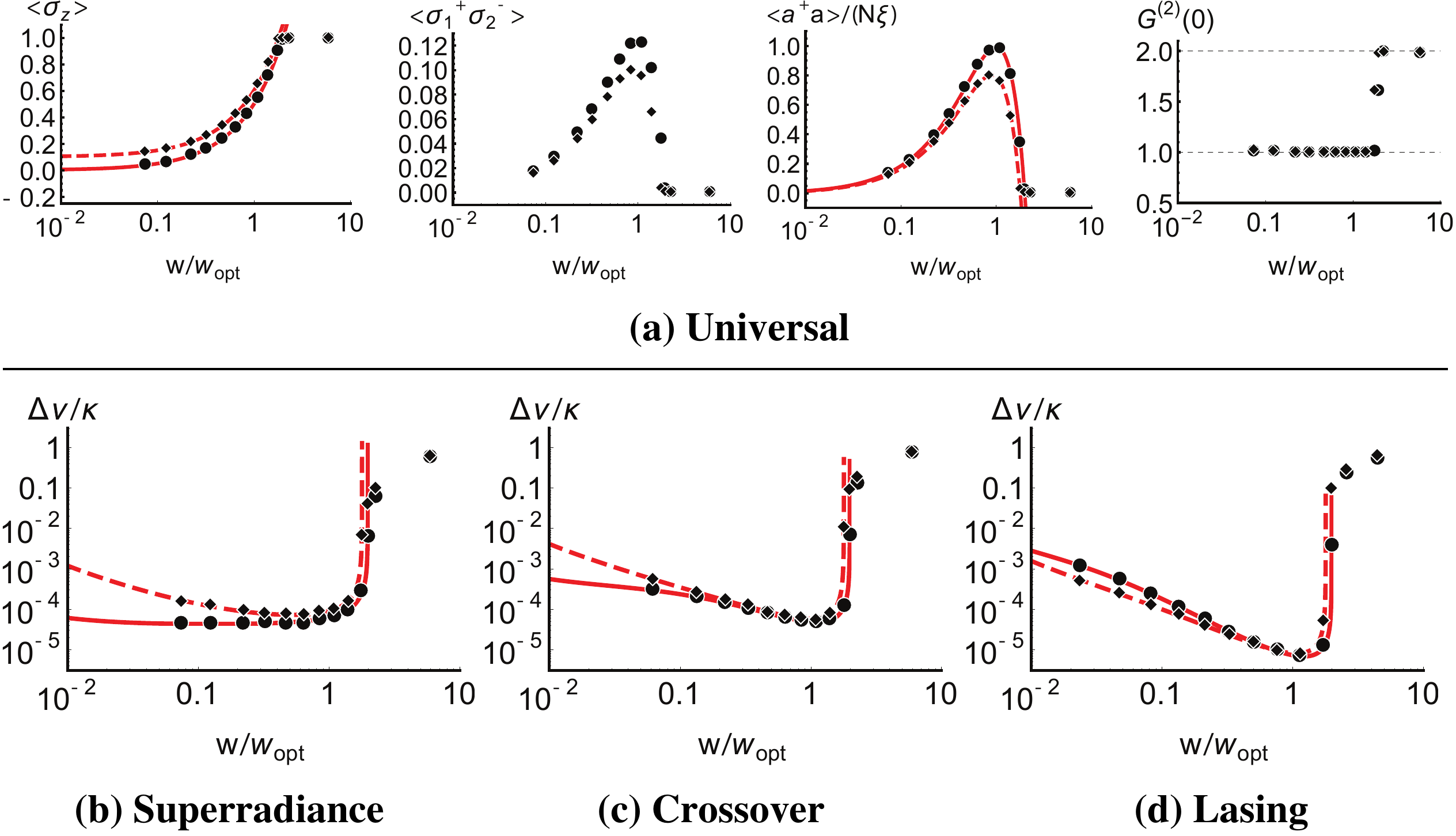}
\end{center}
		\vspace{-5mm}
\caption{(Color online) Solutions using the various methods in the
superradiance ($\xi=0.1$), crossover ($\xi=1$), and lasing ($\xi=10$)
regions for $N=10^4$ and $\frac{\Omega^2}{\kappa \gamma}=0.1$. For
$1/T_2=0$, the analytic Langevin (phase diffusion and mean field)
solutions are shown in solid red (solid light gray), and the
\textit{c}-number Langevin simulation results are shown by black circles.
For $1/T_2=\frac{1}{5} w_{\rm opt}$, the analytic Langevin solutions are
shown in dashed red (dashed light gray), and the \textit{c}-number
Langevin simulation results are shown by black diamonds. (a) All
observables considered except linewidth $\Delta \nu$ show universal
behavior in the superradiance, crossover, and lasing regions, after
appropriate scaling.  (b) $\Delta \nu / \kappa$ in the superradiance
region (c) $\Delta \nu / \kappa$ in the crossover region, (d) $\Delta
\nu / \kappa$ in the lasing region.}
\label{N10000Comparison}
\end{figure*}

\subsection{Many-Atom Characteristics of the Crossover}

Now that the accuracy of the semi-classical \textit{c}-number Langevin
method has been established by comparison with the exact $\mathrm{SU}(4)$ theory,
we study the semiclassical approach in more experimentally realistic
systems with $N=10^4$. The results of these simulations are shown in
Fig.~\ref{N10000Comparison}.  We also include the mean-field Langevin
theory, and the phase diffusion method for the linewidth (see
Appendix~\ref{HakenAppendix}). We consider both the case of vanishing
inhomogeneous broadening, $1/T_2=0$, as well as $1/T_2=w_{\rm opt}/5$.

As seen in Fig.~\ref{N10000Comparison} (a), the inversion
$\left<\hat{\sigma}^{z}\right>$, the correlation between atoms
$\left<\hat{\sigma}_{1}^{+} \hat{\sigma}_{2}^{-}\right>$, the
intracavity photon number $\left<\hat{a}^{\dagger}\hat{a}\right>$, and
the intensity correlation function
$G^{(2)}(0)$~\cite{meystre2007elements} all show universal behavior in
the superradiance, crossover, and lasing regimes after appropriate
scaling.  The calculations for different values of $T_2$ show that
throughout the crossover the system is insensitive to atomic dephasing
provided that the repumping rate is larger than the dephasing rate.

The linewidth $\Delta \nu$ does not show universal behavior in the
superradiance, crossover, and lasing regimes. As seen in
Fig.~\ref{N10000Comparison} (b), in the superradiance region, when
$1/T_2=0$, $\Delta \nu$ is constant in the region of $w<w_{\rm opt}$.
In contrast, the linewidth in the lasing regime, shown in
Fig.\ref{N10000Comparison} (d), linearly decreases as $w$ increases
towards $w_{\rm opt}$. This is the typical Schawlow-Townes behavior of
the laser. In the crossover region, shown in
Fig.~\ref{N10000Comparison} (c), we see that for $w\ll w_{\rm opt}$,
$\Delta \nu$ is constant, and as $w$ approaches $w_{\rm opt}$,
$\Delta \nu$ starts to linearly decrease in a similar manner to its
behavior in the lasing regime. This is a consequence of the property
that a system in the crossover region displays characteristics of both
superradiance and lasing.

When $1/T_2$ is increased to $1/T_2=w_{\rm opt}/5$,
Fig.~\ref{N10000Comparison} (b) shows that $\Delta \nu$ increases for
$w\ll w_{\rm opt}$, but is not significantly affected by $1/T_2$ as $w$
approaches $w_{\rm opt}$. In the crossover region, seen in
Fig.~\ref{N10000Comparison} (c), the behavior is similar. In the lasing
regime, as shown in Fig.~\ref{N10000Comparison}~(d), a qualitatively
different result is observed. The linewidth decreases in the region
slightly below $w=w_{\rm opt}$ for $1/T_2=w_{\rm opt}/5$ when compared
to the $1/T_2=0$ case. This reduction has also been observed for smaller
atom numbers using the exact $\mathrm{SU}(4)$ method, so this
interesting and counterintuitive result is not a consequence of the
failure of the semiclassical approximation.

\begin{figure}
\begin{center}
\includegraphics[scale =0.55] {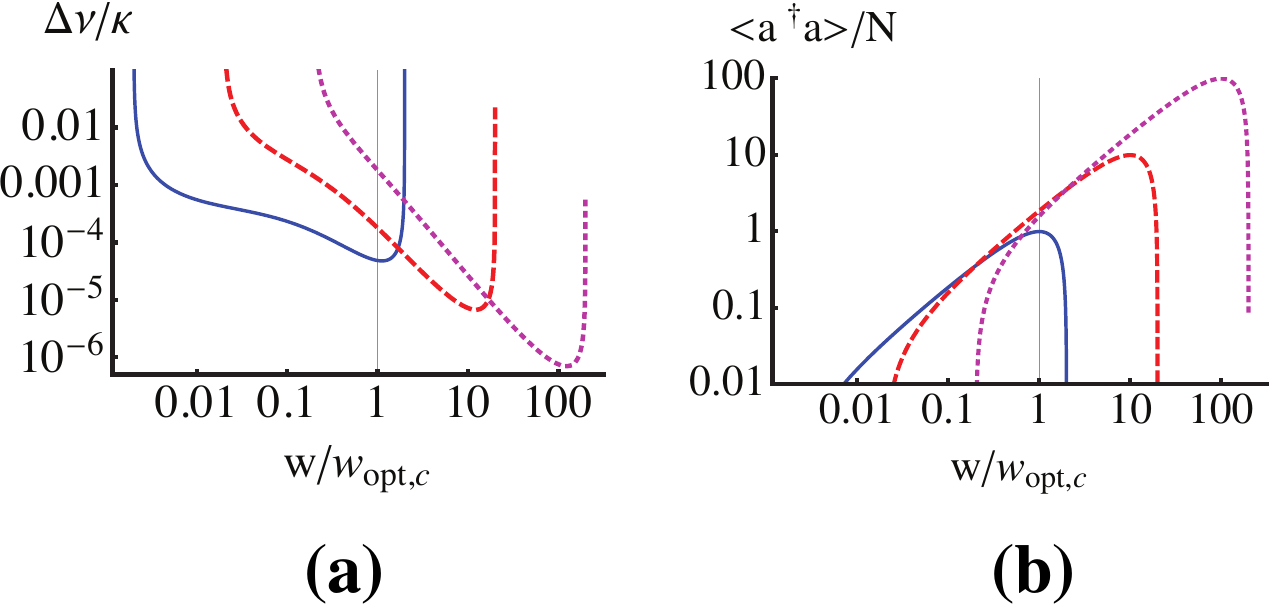}
\end{center}
		\vspace{-5mm}
\caption{(Color online) Comparison of linewidth (a) and intra-cavity
intensity (b) for a system in the crossover regime ($\xi=1$) shown by
the solid blue line, in the lasing regime ($\xi=10$) shown by the dashed
red line, and far into the lasing regime ($\xi=100$) shown by the dotted
magenta line.  Linewidths and intensities were obtained using the
analytic (phase diffusion and mean field) Langevin model. $w_{\rm
opt,c}$ is the optimum $w$ value in the crossover region. For all
systems, $N=10^4$ and $\frac{\Omega^2}{\kappa \gamma}=0.1$.}
\label{LWadaComparison}
\end{figure}

Fig.~\ref{LWadaComparison} shows the potential advantages of operating
in the crossover regime, rather than in the regime of a conventional
laser. Most conventional lasers are limited by available repump power,
and cannot operate at the repump rate that would achieve the greatest
output power and smallest spectral linewidth. It is therefore
interesting to compare crossover and lasing systems operating at the
same absolute repump rate.  As seen in Fig.~\ref{LWadaComparison} (a),
for the same pump rate $w$, a system in the crossover region can
operate with $w=w_{\rm opt}$, whereas for a lasing system, that repump
rate would imply $w\ll w_{\rm opt}$. For this same absolute repump
rate, the crossover system may obtain a linewidth that is orders of
magnitude smaller than the linewidth of the system in the lasing
parameter regime.  Fig.~\ref{LWadaComparison} (b) shows that this
improvement in linewidth can be achieved without paying the penalty of
a greatly reduced output intensity. At this $w$, the output
intensities of the two systems are comparable. 

\subsection{Robustness Against Frequency Shifts}

The sensitivity to frequency shifts is another figure of merit of an
ultrastable light source, especially in the context of precision
measurements.  The linewidth of the emitted light as discussed so far
assumes a perfectly stable cavity frequency and atomic transition
frequency.  However, in the real world these frequencies can vary.  For
example, thermal fluctuations of the cavity mirrors or of the dielectric
coatings on the mirror surfaces can cause fluctuations of the cavity
resonance frequency.  Fluctuating electromagnetic fields, either through
stray fields or through the variation of the black-body radiation that
can arise due to temperature variations, can cause atomic level shifts.
In this Section, we explore the robustness of the ultrastable light
sources to these imperfections.

\begin{figure}
\begin{center}
\includegraphics[scale =0.6] {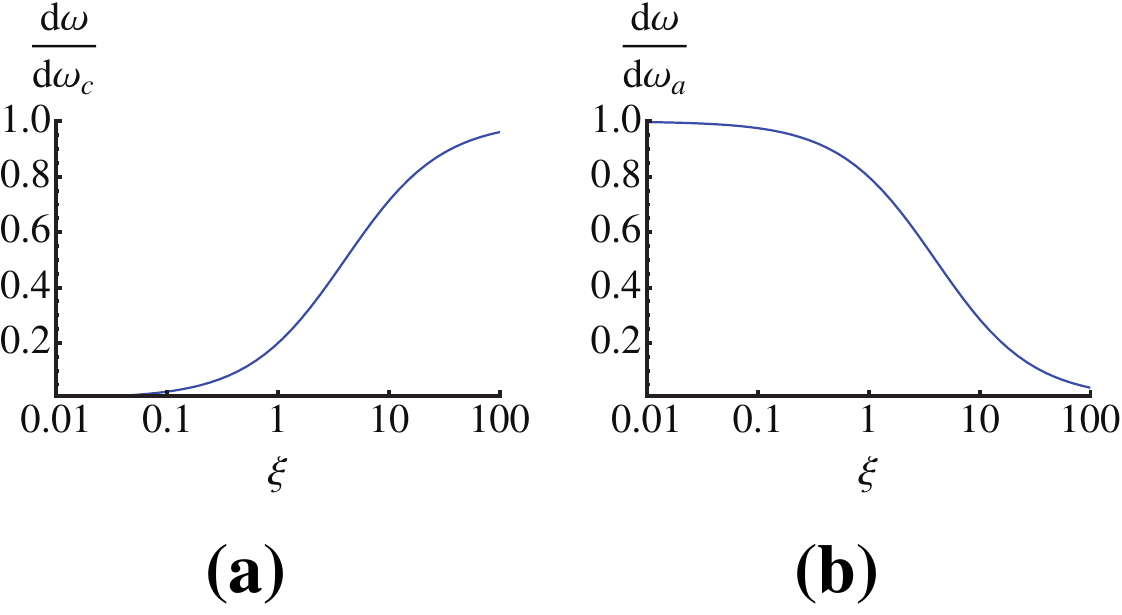}
\end{center}
\vspace{-5mm}
\caption{(Color online) Instability in the atom-cavity system frequency
$\omega$ with respect to (a) the cavity frequency $\omega_c$ and (b)
atomic frequency $\omega_a$ as a function of the crossover parameter
$\xi$.}
\label{CavityInstability}
\end{figure}

We illustrate in Fig.~\ref{CavityInstability} the sensitivity of the
line-center of the spectrum of the output light with respect to both
the cavity resonance frequency and the atomic resonance frequency as a
function of the crossover parameter. The sensitivity is characterized
by the derivatives of the line-center frequency with respect to
$\omega_c$ and $\omega_a$.  On the superradiance side of the
crossover, $\xi \ll 1$, we observe that the system is sensitive to
fluctuations of the atomic resonance frequency but robust against
fluctuations of the cavity resonance. The situation is reversed on the
lasing side. As a consequence, by continuously varying the crossover
parameter, one has control of the relative importance of cavity
frequency and atom level noise to the spectrum of the ultrastable
light that is produced.

\section{Conclusion}

In this paper we have theoretically studied the continuous crossover
from steady state superradiance to lasing.  We have defined a
dimensionless crossover parameter that characterizes the regime as the
ratio of the maximum intracavity photon number to the atom number. We
showed that this encapsulates the relative importance of stimulated
emission and collective superradiance. We developed a semiclassical
method based on \textit{c}-number Langevin equations and verified the
accuracy of this method by comparison with exact numerical solutions.

We have systematically investigated a range of important observables;
the output intensity, the linewidth of the emitted light, intensity
correlation functions, and the sensitivity to perturbations of the
cavity and atomic resonance frequencies.  We find that when the repump
rate is constrained, a system in the crossover regime may operate with a
much smaller intrinsic linewidth and be less sensitive to cavity pulling
than a comparable system operating as a conventional laser.

\begin{acknowledgments}
  This work was supported by the DARPA ATN program through grant
  number W911NF-16-1-0576 through ARO. The views and conclusions
  contained in this document are those of the authors and should not
  be interpreted as representing the official policies, either
  expressed or implied, of the U.S. Government. The U.S. Government is
  authorized to reproduce and distribute reprints for Government
  purposes notwithstanding any copyright notation herein. D.T. and
  D.M acknowledge support from the National Science Foundation under
  Grants PHY-1521080, PHY-1404263 and PHY-1125844.
\end{acknowledgments}

\appendix

\section{$\mathrm{SU}(4)$ Simulation of the Quantum Master Equation}
\label{Su4Appendix}

In this appendix we provide a short summary of the method we are using
for the direct numerical simulation of the open quantum.  Although the
general aspects of the formalism are detailed in
reference~\cite{PhysRevA.87.062101}, here we extend the $\mathrm{SU}(4)$
method to allow us to simulate systems with a moderate number of photons
and atoms.  We do this by unraveling the quantum master equation into
quantum trajectories---a standard method in quantum optics.  The unusual
feature here is that the unraveling is performed in Louville space
rather than in Hilbert space, because it is in Louville space that the
$\mathrm{SU}(4)$ method operates.

The key insight is to exploit the invariance of the master
equation~(\ref{ME1Crossover}) under particle exchange.  This permutation
symmetry allows us to write the equations of motion in terms of
generators of the $\mathrm{SU}(4)$ group.  The Hamiltonian becomes
\begin{eqnarray}
  &&\frac{1}{i\hbar}[H,\hat{\rho}]=
  -2i \omega_a \Sigma_3\hat{\rho} -i\omega_c [ \hat{a}^{\dagger}\hat{a}, \hat{\rho}]
  \nonumber
  \\
  &&-i\Omega \left[a(\mathcal{M}_++\mathcal{N}_+)\hat{\rho}+a^\dagger
    (\mathcal{M}_-+\mathcal{N}_-)\hat{\rho}\right]\nonumber\\
  &&\quad{}+i\Omega\left[(\mathcal{U}_++\mathcal{V}_+)\hat{\rho} a^\dagger
    +(\mathcal{U}_-+\mathcal{V}_-)\hat{\rho} a\right]\,,
\end{eqnarray}
and the dissipation terms become
\begin{equation}
  \frac12\sum_{j=1}^N(
   2\sigma_j^{\mp}\hat{\rho}\sigma_j^{\pm}-\sigma_j^{\pm} \sigma_j^{\mp}\hat{\rho}-
   \hat{\rho} \sigma_j^{\pm}\sigma_j^{\mp}
  )=-\frac{N}{2}\mp
  \mathcal{Q}_3+\mathcal{Q}_{\mp}\,,
\end{equation}
for the population changing terms, and
\begin{equation}
\sum_{j=1}^N(\sigma_j^{z}\hat{\rho}\sigma_j^{z}-\hat{\rho})=4\mathcal{M}_3-2
  \mathcal{Q}_3-2\Sigma_3-N\,,
  \label{ham}
\end{equation}
for the dephasing term.  In these equations, $\mathcal{Q}_{\pm}$,
$\mathcal{M}_{\pm}$, $\mathcal{N}_{\pm}$, $\mathcal{U}_{\pm}$,
$\mathcal{V}_{\pm}$, $\mathcal{Q}_3$, $\mathcal{M}_3$, and $\Sigma_3$
are superoperators~\cite{PhysRevA.87.062101}.

We expand the density matrix in terms of the fully symmetrical multiplet
$P_{q,q_3,\sigma_3}$~\cite{PhysRevA.87.062101} of the $\mathrm{SU}(4)$ group,
\begin{equation}\label{ex}
  \hat{\rho}=\sum_{q,q_3,\sigma_3,m,n} C_{q,q_3,\sigma_3}^{m,n}
  P_{q,q_3,\sigma_3}\bigl|m\bigr>\bigl<n\bigr|\,,
\end{equation}
where $C_{q,q_3,\sigma_3}^{m,n}$ are complex coefficients, and
$|n\rangle$ is the photon Fock state. Note that the total number of
states in the fully symmetrical multiplet is $(N+1)(N+2)(N+3)/6$,
which reduces the exponential scaling of the problem to cubic in $N$.

However, the dimension of the density matrix grows as the square of the
photon number.  This would impose great difficulties in numerical
simulations of the laser region due to the large number of photons.  To
overcome this difficulty, we unravel the master equation into
Monte-Carlo trajectories in Liouville space enabling us to eliminate the
photon basis from the simulation.  The essential idea behind the method
is that we are able to deduce the photon state by keeping track of the
total number of quanta $N_q$ in the system.

The quantum Monte Carlo method decomposes the density operator evolution
into a set of quantum trajectories where, between applications of random
jumps into random channels, the system evolves under an effective
Hamiltonian~\cite{Dalibard92,Dum92,Knight98}.  The random jumps are
chosen with probabilities such that the correct density operator
evolution is obtained when an average is taken over trajectories.  To
construct a single trajectory, we first need to identify the jump
operators. In our problem, there are four decay channels: repumping,
spontaneous emission, dephasing, and cavity decay. The corresponding
jump operators $\mathcal{J}_i$ are
\begin{eqnarray}
\mathcal{J}_1\hat{\rho}&=&
w\sum_{j=1}^N(\sigma_j^+\hat{\rho}\sigma_j^-)=w\mathcal{Q}_{+}\hat{\rho}\,,
\nonumber\\
\mathcal{J}_2\hat{\rho}&=&
\gamma\sum_{j=1}^N(\sigma_j^-\hat{\rho}\sigma_j^+)=
\gamma \mathcal{Q}_{-}\hat{\rho}\,,\nonumber\\
\mathcal{J}_3\hat{\rho}&=&
\frac{1}{2T_2}\sum_{j=1}^N(\sigma_j^{z}\hat{\rho}\sigma_j^{z})
\nonumber\\
&=&\frac{1}{2T_2}(4\mathcal{M}_3-2  \mathcal{Q}_3-2\Sigma_3)\hat{\rho}\,,
\nonumber\\
\mathcal{J}_4\hat{\rho}&=&\kappa a\hat{\rho} a^{\dagger}\,.
\label{jumpo}
\end{eqnarray}
When a repumping quantum jump occurs, $N_q$ increases by one.  When a
spontaneous emission or a cavity-decay quantum jump happens, $N_q$
decreases by one.  The dephasing quantum jumps leave $N_q$ unchanged.
Therefore, during the evolution of a single trajectory, $N_q$ is
uniquely determined at every time step by keeping track of the numbers
of jumps of the different types. With knowledge of $N_q$, the photon
number does not need to be treated as an independent variable but is
determined from the number of excited atoms.  In
Ref.~\cite{PhysRevA.87.062101} we have shown that
\begin{equation}
\begin{split}
  \hat{J}_z P_{q,q_3,\sigma_3}^{(\mathrm{s})}&=
  (q_3+\sigma_3)P_{q,q_3,\sigma_3}^{(\mathrm{s})}\;,\\
  P_{q,q_3,\sigma_3}^{(\mathrm{s})}\hat{J}_z&=
  (q_3-\sigma_3)P_{q,q_3,\sigma_3}^{(\mathrm{s})}\;,
\end{split}
\end{equation}
where $\hat{J}_z=\sum_{j=1}^N\sigma_j^{z}/2$ is the collective spin
operator.  Therefore, the atomic state for a particular fully
symmetrical atomic basis state in terms of the number of excited atoms
is $|q_3+\sigma_3+N/2\rangle\langle q_3-\sigma_3+N/2|$.  And thus the
corresponding photon state is
\begin{equation}
\bigl|m\bigr>\bigl<n\bigr|=
\bigl|N_q-(q_3+\sigma_3+N/2)\bigr>
\bigl<N_q-(q_3-\sigma_3+N/2)\bigr|.
\end{equation}

The simulation of jump times and decay channels is completely analogous
to the wave-function Monte Carlo method. The effective evolution of the
system is governed by the master equation excluding the above jump
operators. As a result, under the effective evolution, the trace of the
density operator is no longer conserved, but decreases as a function of
time.  This is analogous to the decay of the norm of the wavefunction in
the wave function Monte Carlo method. A jump occurs when the trace of
the density operator is less than a random number uniformly distributed
in the interval $[0,1]$. When a decay occurs we stochastically determine
the channel $i$ into which the system decays according to the
probability distribution,
\begin{equation}
\mathcal{P}_i^{\mathrm{jump}}=
\frac{\mathrm{Tr}[\mathcal{J}_i\hat{\rho}]}
{\sum_{k=1}^4 \mathrm{Tr}[\mathcal{J}_k\hat{\rho}]}\;.
\end{equation}

Finally, in order to get the density operator at each time step, an
ensemble average of many quantum trajectories is required. Then, various
observables can be calculated according to
Ref.~\cite{PhysRevA.87.062101}. It is also worth noting that if one is
only interested in the steady state density operator, a time average in
the steady state can be applied instead of the ensemble average.

\section{Phase Diffusion Linewidth}
\label{HakenAppendix}

In this appendix we derive a closed form expression for the linewidth
based on a phase diffusion model.  Our analysis closely follows the
derivation in~\cite{HakenLaser, HakenLaserBook}.

Differentiating Eq.~(\ref{La}) with respect to time and substituting 
Eqs.~(\ref{La})~--~(\ref{Lsm}) we obtain
\begin{equation}
\ddot{\hat{a}} =
-\frac{1}{2} (\kappa+\Gamma)  \dot{\hat{a}} -
\frac{\kappa \Gamma}{4}\hat{a}  +
\frac{N \Omega^2 }{4} \hat{a} \hat{S}^z +\hat{F}\;,
\label{addeq}
\end{equation}
where
\begin{eqnarray}
\hat{S}^z &=&
\int_0^t dt^{\prime} e^{-(w+\gamma)(t-t^{\prime})}
\bigg( (w+\gamma) + \hat{F}^z
\nonumber
\\
&&\hspace{-3mm}-\frac{2}{N} \left( \frac{d}{dt} (\hat{a}^{\dagger} \hat{a}) +
\kappa \hat{a}^{\dagger} \hat{a} -\hat{a}^{\dagger} \hat{F}^a -
\hat{F}^{a^{\dagger}} \hat{a} \right) \bigg)\;,
\end{eqnarray}
and
\begin{equation}
\hat{F} = \frac{\Gamma}{2} \hat{F}^a-
\frac{i N \Omega}{2} \hat{F}^-+\dot{\hat{F}}^a\;.
\end{equation}

The annihilation operator $\hat{a}$ is decomposed according to
\begin{equation}
\hat{a}= (a_0 + \hat{\rho}) e^{i\hat{\phi}}\;.
\label{adecomp}
\end{equation}
Above threshold, amplitude fluctuations are small so that $\hat\rho$
can be neglected.  We then obtain for the two time correlation function
of the field amplitude
\begin{equation}
\left< \hat{a}^{\dagger}(t) \hat{a}(0) \right> =
a_0^2 \left< e^{i(\hat{\phi}(t) - \hat{\phi}(0))} \right>\;.
\end{equation}

After substituting Eq.~(\ref{adecomp}) into Eq.~(\ref{addeq}), we take
the imaginary part to first order in products of operators, and find,
\begin{equation}
\ddot{\hat{\phi}} =
-\frac{1}{2}(\kappa+\Gamma) \dot{\hat{\phi}} +
\frac{1}{a_0} \text{Im} [\hat{F}]\;,
\label{phieq1}
\end{equation}
where a factor of $e^{-i\phi}$ has been absorbed into $\hat{F}$.
Equation~(\ref{phieq1}) is then integrated, assuming that
$(\kappa+\Gamma)$ is large, to arrive at
\begin{equation}
\hat{\phi}(t) - \hat{\phi}(0) =
\frac{2}{a_0 (\kappa+\Gamma)}
\int_0^t dt^{\prime} \text{Im}
\left[ \frac{\Gamma}{2} \hat{F}^a-\frac{i N \Omega}{2} \hat{F}^-\right]\;.
\label{phieq2}
\end{equation}
Since $ \hat{F}^a$ and $\hat{F}^-$ are Gaussian, we can use
\begin{equation}
\left< e^{i(\hat{\phi}(t) - \hat{\phi}(0))} \right> =
e^{-\frac{1}{2}\left< {( \hat{\phi}(t) - \hat{\phi}(0) )}^2 \right>}\;.
\end{equation}
Therefore, we use Eq.~(\ref{phieq2}), along with Eqs.~(\ref{OpNoise1})
to find
\begin{equation}
\left<{(\hat{\phi}(t) - \hat{\phi}(0))}^2 \right> =
\frac{(C+1)}{2(Cd_0-1)} \frac{\Gamma}{(w+\gamma)}
\frac{\Omega^2 \kappa}{{(\kappa+\Gamma)}^2} t\;,
\end{equation}
so that the linewidth $\Delta \nu$ given by
\begin{equation}
\Delta \nu =
\frac{(C+1)}{2(Cd_0-1)} \frac{\Gamma}{(w+\gamma)}
\frac{\Omega^2 \kappa}{{(\kappa+\Gamma)}^2}\;.
\label{LWHaken}
\end{equation}

\bibliography{CrossoverPaper}

\end{document}